\documentclass{iaus}
\usepackage{graphicx}

\def\Alfven{Alfv\'{e}n~}
\def\Alfvenic{Alfv\'{e}nic~}

\title[Alfv\'{e}n wave-driven winds] 
{Evolution of Alfv\'{e}n wave-driven \\
solar winds to red giants}

\author[T.K. Suzuki]   
{Takeru K. Suzuki
}

\affiliation{
School of Arts and Sciences, University of Tokyo,
Komaba, Meguro, Tokyo, Japan 153-8902\\email: {\tt
stakeru@ea.c.u-tokyo.ac.jp}}

\pubyear{2008}
\volume{247}  
\pagerange{xxx--xxx}
\date{?? and in revised form ??}
 \jname{Waves \& Oscillations in The Solar
Atmosphere: \\Heating and Magneto-Seismology} \editors{R.
Erd\'{e}lyi \& C.~A. Mendoza-Brice$\tilde{n}$o,
eds.}
\begin{document}

\maketitle

\begin{abstract}
In this talk we introduce our recent results of global 1D MHD simulations for 
the acceleration of solar and stellar winds.
We impose transverse photospheric motions corresponding to the granulations, 
which generate outgoing Alfv\'{e}n waves.
The \Alfven waves effectively dissipate by 3-wave coupling and direct mode
conversion to compressive waves in density-stratified atmosphere. 
We show that the coronal heating and the solar wind acceleration in the 
open magnetic field regions are natural consequence of the footpoint 
fluctuations of the magnetic fields at the surface (photosphere).  
We also discuss winds from red giant stars driven by \Alfven waves, focusing 
on different aspects from the solar wind. 
We show that red giants wind are highly structured with intermittent 
magnetized hot bubbles embedded in cool chromospheric material.

\end{abstract}

\firstsection 
\section{Introduction}
The \Alfven wave, generated by the granulations or other surface 
activities, is a promising candidate operating in the heating and 
acceleration of solar winds from coronal holes. 
It can travel a long distance so that the dissipation plays a role 
in the heating of the solar wind plasma as well as the lower coronal plasma, 
in contrast to other processes, such as magnetic reconnection 
events and compressive waves, the heating of which probably concentrates 
at lower altitude. 

While high-frequency ioncyclotron waves are recently highlighted for 
the preferential heating of minor heavy ions (Axford \& McKenzie 1997)
the protons,  
which compose the main part of the plasma, are supposed to be mainly heated by 
low-frequency ($\lesssim 0.1$Hz) components in the MHD regime.  
because (1) the low-frequency wave is expected 
to have more power, and (2) the resonance frequency of 
the proton is higher than those of heavier ions so that the energy of the 
ioncyclotron wave is in advance absorbed by heavy ions (Cranmer 2000).
In this paper, we focus on roles of such low frequency \Alfven waves. 

When considering low-frequency \Alfven waves in solar and stellar 
atmospheres, the stratification of the density due to the gravity 
is quite important because the variation scale of the density, 
and accordingly \Alfven speed, is comparable to or shorter than the 
wavelengths; the WKB approximation is no longer applicable. 
Also, the amplitude 
is amplified because of the decrease of the density so that waves easily 
become nonlinear. 
Recently, we have extensively studied the heating and acceleration of 
solar and stellar winds by self-consistent MHD simulations from the 
photosphere to sufficiently outer region  
(Suzuki \& Inutsuka 2005; 2006; hereafter SI06; Suzuki 2007). 
We review these works in this contribution talk.   

\section{Simulation}
We consider 1D open flux tubes which are super-radially open, 
measured by heliocentric distance, $r$. 
The simulation regions are from the photosphere 
(density, $\rho = 10^{-7}$g cm$^{-3}$, for the Sun) to several stellar 
radii. 
Radial field strength, $B_r$, 
is given by conservation of magnetic flux as $B_r r^2 f(r) = {\rm const.}$, 
where $f(r)$ is a super-radial expansion factor(see SI06 
for detail). 

We input the transverse fluctuations of the field line by the 
granulations at the photosphere, which excite \Alfven waves. 
In this paper we only show results of linearly polarized perturbations 
with power spectrum proportional to $1/\nu$, where $\nu$ is frequency (for 
circularly polarized fluctuations with different spectra, see SI06).
Amplitude, $\langle dv_{\perp,0} \rangle$, at the photosphere is chosen to 
be compatible with  
the observed photospheric velocity amplitude $\sim 1$km s$^{-1}$ 
(Holweger 1978).
At the outer boundaries, non-reflecting condition is imposed for all the MHD 
waves, which enables us to carry out 
simulations for a long time until quasi-steady state solutions are obtained  
without unphysical wave reflection. 

We dynamically treat the propagation and dissipation of the waves and the 
heating and acceleration of the plasma by solving ideal MHD equations.  
In the energy equation we take into account radiative cooling and Spitzer 
thermal conduction (SI06).  
We adopt the second-order MHD-Godunov-MOCCT scheme (Sano \& Inutsuka 2008 
in preparation) 
to update the physical quantities. 
We initially set static atmosphere with a temperature $T=10^4$K to see 
whether the atmosphere is heated up to coronal temperature and accelerated 
to accomplish the transonic flow. 
At $t=0$ we start the inject of the transverse fluctuations from the 
photosphere and continue the simulations until the quasi-steady states 
are achieved.  
   
\section{Results}

\subsection{Fast Solar Wind}

\begin{figure}[h!]
   \hskip2cm
   \rotatebox{0}{\resizebox{12cm}{!}{\includegraphics{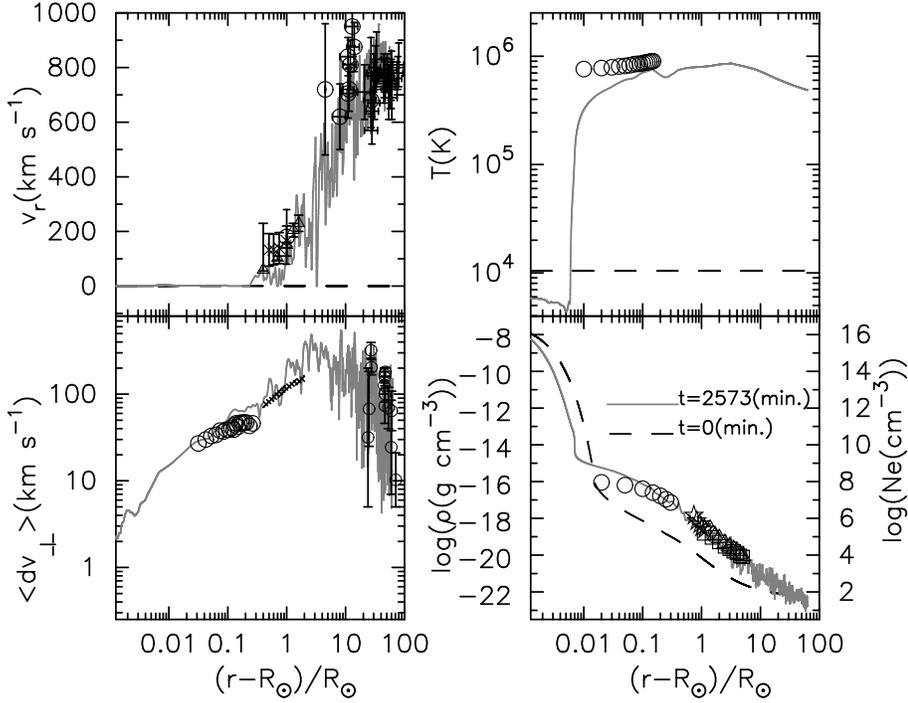}}}
  \caption{Results of fast solar wind mode with observations in polar regions. 
Outflow speed, $v_r$(km s$^{-1}$) (top-left), temperature, $T$(K) (top-right), 
density in logarithmic scale, $\log(\rho({\rm g\;cm^{-3}}))$ (bottom-right), 
and rms transverse amplitude, $\langle dv_{\perp} \rangle$(km s$^{-1}$) 
(bottom-left) are plotted. 
Observational data in the third panel are electron density, 
$\log(N_e({\rm cm^{-3}}))$ which is to be referred to the right axis. 
Dashed lines indicate the initial conditions and solid lines 
are the results at $t=2573$ minutes. In the bottom panel, the initial 
value ($\langle dv_{\perp} \rangle=0$) dose not appear. 
The observational data in the inner region ($<6R_{\odot}$) are from SOHO 
(Teriaca et al.2003; Zangrilli et al.2002; Fludra et al.1999; 
Wilhelm et al.1998; Lamy et al.1997; Banergee et al.1998; Esser et al.1999)
and those in 
the outer region are from interplanetary scintillation measurements 
(Grall et al.1996; Habbal et al.1994; Kojima et al.2004; Canals et al.2002).
}
  \label{fig:obs}
\end{figure}

Figure \ref{fig:obs} plots the initial condition (dashed lines) and 
the results after the quasi-steady state condition is achieved at $t=2573$ 
minutes (solid lines), 
compared with recent observations of fast solar winds. 
We set the transverse fluctuation, $\langle dv_{\perp,0}
\rangle = 0.7$km s$^{-1}$, and field strength, $B_{r,0} = 161$G at 
the photosphere, and the total superradial expansion factor, 
$f_{\rm tot} = 75$.  
Figure \ref{fig:obs} shows that the initially cool and static atmosphere 
is effectively heated and accelerated by the dissipation of the \Alfven waves. 
The sharp transition region which divides the cool chromosphere with 
$T\sim 10^4$K and  
the hot corona with $T\sim 10^6$K is formed owing to a thermally unstable 
region around $T\sim 10^5$K in the radiative cooling function 
(Landini \& Monsignori-Fossi 1990). 
The hot corona streams out as the transonic solar wind. 
The simulation naturally explains the observed trend quite well. 

\begin{figure}[h]
   \hskip2cm
   \rotatebox{0}{\resizebox{7cm}{!}{\includegraphics{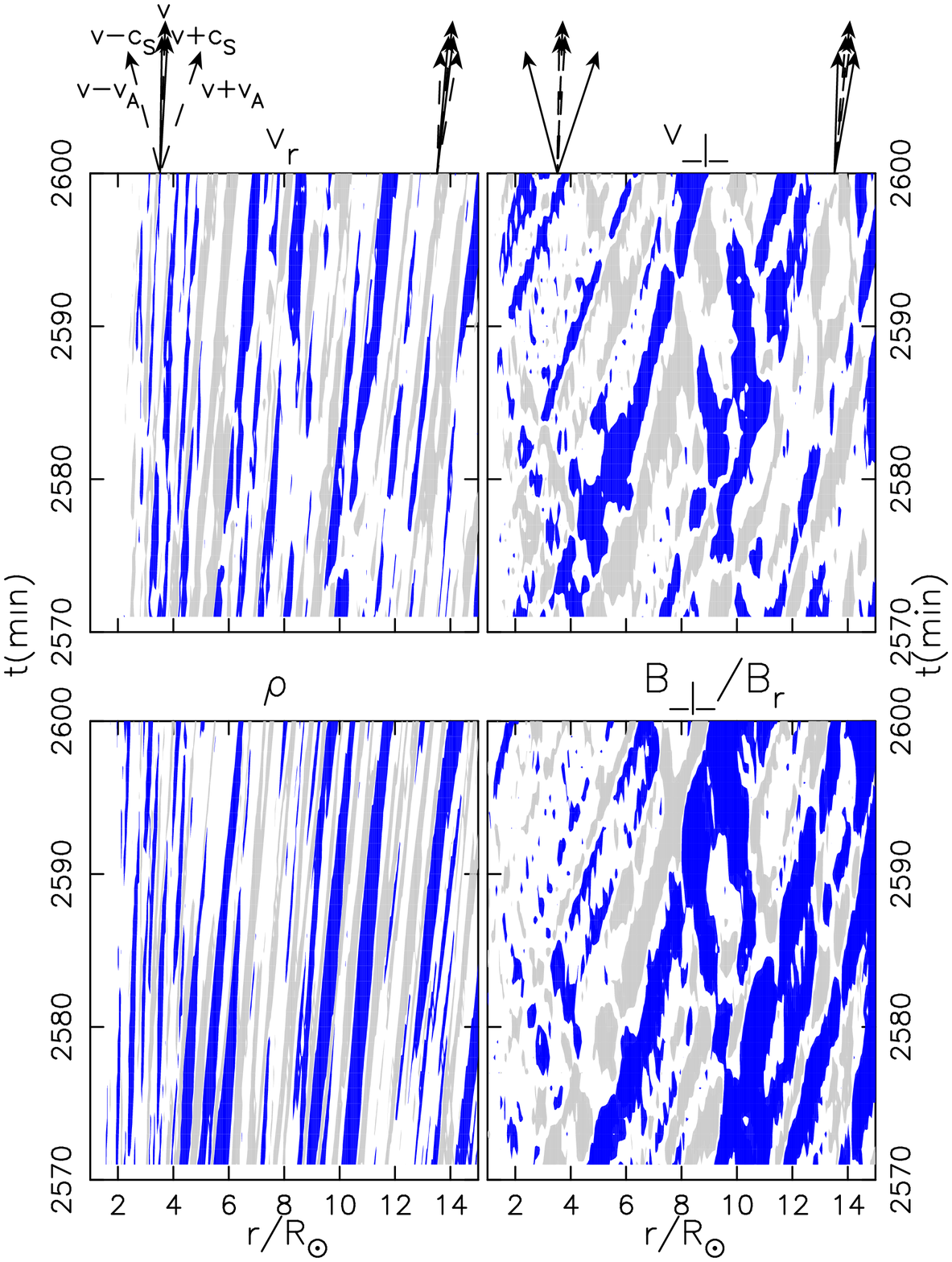}}}
  \caption{$r-t$ diagrams for $v_r$ (upper-left), $\rho$ (lower-left), 
$v_{\perp}$ (upper-right), and $B_{\perp}/B_r$ (lower-right.) 
The horizontal axises cover from $R_{\odot}$ to $15R_{\odot}$, and 
the vertical axises cover from $t=2570$ minutes to $2600$ minutes.  
Dark(blue) and light shaded regions indicate positive and negative amplitudes 
which exceed certain thresholds. The thresholds are $d v_r 
=\pm 96$km/s for $v_r$, $d\rho /\rho=\pm0.25$ for $\rho$, 
$v_{\perp}=\pm 180$km/s for $v_{\perp}$, and $B_{\perp}/B_r=\pm 
0.16$ for $B_{\perp}/B_r$, where $d \rho$ and $d v_r$ are 
differences from the averaged $\rho$ and $v_r$. 
Arrows on the top panels indicate characteristics of \Alfven, slow MHD 
and entropy waves at the respective locations. }
  \label{fig:tm-dis}
\end{figure}

The heating and acceleration of the solar wind plasma in inner heliosphere 
is done by the dissipation of \Alfven waves. Here we inspect waves  
in more detail. 
Figure \ref{fig:tm-dis} 
presents contours of amplitude of $v_r$, $\rho$, $v_{\perp}$, and 
$B_{\perp}/B_r$ in $R_{\odot} \le r \le 15 R_{\odot}$ 
from $t=2570$ min. to $2600$ min. 
Red (blue) shaded regions denote positive (negative) amplitude. Above 
the panels, we indicate the directions of the local 5 characteristics, two 
\Alfven, two slow, and one entropy waves at the respective positions. 
Note that the fast MHD and \Alfven 
modes degenerate in our case (wave vector and underlying magnetic field are 
in the same direction), so we simply call the innermost and outermost waves 
\Alfven modes.  
In our simple 1D geometry, $v_r$ and $\rho$ trace the slow modes 
which have longitudinal wave components, while $v_{\perp}$ and $B_{\perp}$ 
trace the \Alfven modes which are transverse 

One can clearly see the \Alfven waves in $v_{\perp}$ and $B_{\perp}/B_r$ 
diagrams, which have the same slopes with the \Alfven characteristics shown 
above. 
One can also find the incoming modes propagating from lower-right to 
upper-left as well as the outgoing modes generated from the surface
These incoming waves are generated by the reflection at the `density mirrors'  
of the slow modes.
At intersection points of the outgoing and incoming characteristics 
the non-linear wave-wave interactions take place, which play a role 
in the wave dissipation. 

The slow modes are seen in $v_r$ and $\rho$ diagrams. Although it might 
be difficult to distinguish, the most of the patterns are due to the outgoing 
slow modes\footnote{The phase correlation of the longitudinal slow 
waves is opposite to that of the transverse \Alfven waves. 
The outgoing slow modes 
have the positive correlation between amplitudes of $v_r$ and $\rho$, 
($\delta v_r \delta \rho > 0$), while the incoming modes have the negative 
correlation ($\delta v_r \delta \rho < 0$).}
which are generated from the perturbations of the \Alfven wave 
pressure, $B_{\perp}^2/8\pi$ (Kudoh \& Shibata 1998\& Tsurutani et al. 2002).
These slow waves steepen eventually and lead to the shock dissipation. 

\begin{figure}[t]
   \hskip2cm
   \rotatebox{0}{\resizebox{6.5cm}{!}{\includegraphics{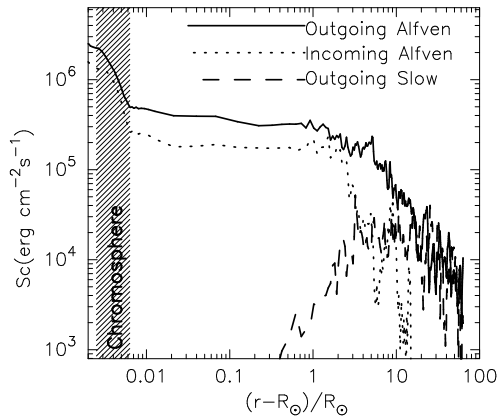}}}
  \caption{$S_c$ of outgoing \Alfven mode (solid), 
incoming \Alfven mode (dotted), and outgoing MHD slow 
mode (dashed) at 
$t=2573$ mins. 
Hatched region indicates the chromosphere and low 
transition region with $T<4\times 10^4$K.}
  \label{fig:wvact}
\end{figure}

Figure \ref{fig:wvact} presents the dissipation of the waves more 
quantitatively. 
It plots the following quantities, 
\begin{equation}
S_c 
=\rho \delta v^2 \frac{(v_r + v_{\rm ph})^2}
{v_{\rm ph}} \frac{r^2 f(r)}{r_c^2 f(r_c)}, 
\end{equation} 
of outgoing \Alfven, incoming \Alfven, and outgoing slow MHD (sound) waves, 
where $\delta v$ and $v_{\rm ph}$ are amplitude and phase speed of each 
wave mode. $S_c$ is an adiabatic constant derived from wave action 
(Jacques 1977) in unit of energy flux. 
For the incoming \Alfven wave, we plot the opposite sign of 
$S_c$ so that it becomes positive in the sub-\Alfvenic region. 
The outgoing and incoming \Alfven waves are decomposed 
by correlation between $v_{\perp}$ and $B_{\perp}$. 
Extraction of the slow wave is also from fluctuating components of $v_r$ and 
$\rho$. 

Figure \ref{fig:wvact} clearly illustrates that 
the outgoing \Alfven waves dissipate quite effectively; 
$S_c$ becomes only $\sim 10^{-3}$ of the initial value 
at the outer boundary. 
A sizable amount is reflected back 
downward below the coronal base ($r-R_{\rm S} < 0.01 
R_{\rm S} $), which is known from the incoming \Alfven wave following 
the outgoing component with slightly smaller level.  
This is because the wave shape is considerably 
deformed owing to the steep density gradient; 
a typical variation scale ($< 10^5$km) 
of the \Alfven speed becomes comparable to or even shorter than 
the wavelength ($=10^4 - 10^6$km). 
Although the energy flux, $\simeq 5\times 10^{5}$erg cm$^{-2}$s$^{-1}$, 
of the outgoing \Alfven waves ($S_c$ in the static region 
is equivalent with the energy flux) which penetrates into the corona is 
only $\simeq 15$\% of the input value,  
it satisfies the requirement for the energy budget in the coronal holes 
(Withbroe \& Noyes 1977).

The processes discussed here are the combination of the direct mode conversion 
to the compressive waves and the parametric decay instability due to 
three-wave (outgoing \Alfven, incoming \Alfven, and outgoing slow waves) 
interactions (Goldstein 1978; Terasawa et al. 1986) 
of the \Alfven waves. 
These processes, which are not generally efficient in homogeneous background, 
become effective by amplification of velocity amplitude in the density 
decreasing atmosphere. 
The \Alfven speed also varies a lot even within one wavelength of \Alfven 
waves with periods of minutes. 
This leads to both variation of the wave pressure in one wavelength  
and partial reflection through the deformation of the wave shape 
(Moore et al.1991).
The density stratification plays a key role in the propagation and dissipation 
of the \Alfven waves. 

\subsection{Evolution to Red Giant Winds}
So far we have focused on the acceleration of solar wind. The same process 
is expected to operate in other types of stars that have 
surface convective layer, such as red giant stars, proto-stars, 
and intermediate and low mass main sequence stars. 
As a demonstration, we have applied our solar wind simulations to red giant 
winds (Suzuki 2007). 

We consider stellar winds from 1$M_{\odot}$ stars in various evolutionary 
stages from main sequence to red giant branch. 
The properties of surface fluctuations (e.g. amplitude and 
spectrum) can be estimated from conditions of surface convection which 
depend on surface gravity and temperature (e.g. Renzini et al.1977; Stein 
et al. 2004). Then, we carry out the simulations of the red giant winds 
in a similar manner to the solar wind simulations. 

\begin{figure}[t]
   \hskip2cm
   \rotatebox{0}{\resizebox{6.7cm}{!}{\includegraphics{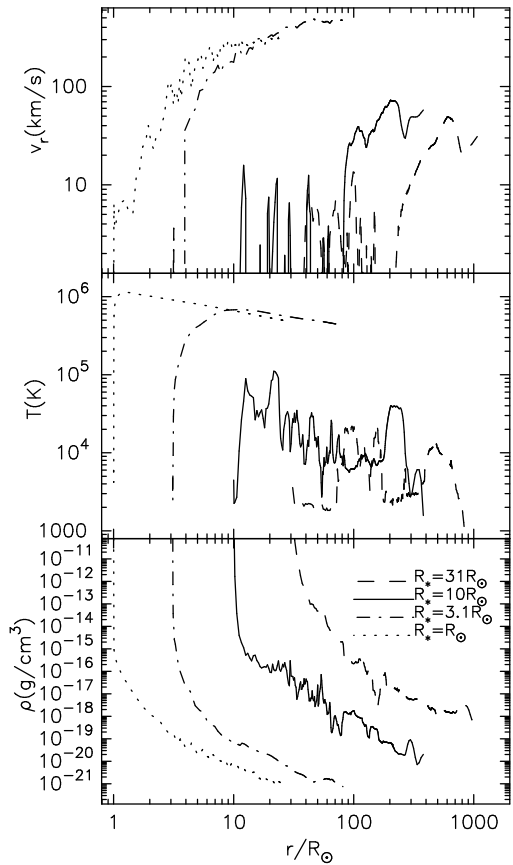}}}
  \caption{Time-averaged stellar wind structure of the $1M_{\odot}$ stars. 
From the top to the 
bottom, radial outflow velocity, $v_r$ (km s$^{-1}$), temperature, $T$ (K),  
and density, $\rho$(g cm$^{-3}$), are plotted. 
The black, blue, green, and red lines are the 
results of stellar radii, $R=R_{\odot}$ (the present Sun), $3.1R_{\odot}$ 
(sub-giant), $10R_{\odot}$ (red giant), and $31R_{\odot}$ (red giant), 
respectively.} 
  \label{fig:rgbwnd}
\end{figure}

Figure \ref{fig:rgbwnd} presents the evolution of stellar winds of 
a $1M_{\odot}$ star from main sequence to red giant stages. 
The middle panel shows that the average temperature drops suddenly 
from $T\simeq 7\times 10^5$K in the sub-giant star (blue) 
to $T\le 10^5$K in the red giant stars, which is consistent with the observed 
``dividing line''(Linsky \& Haisch 1979). 
The main reason of the disappearance of the steady hot coronae is that 
the sound speed ($\approx 150$ km s$^{-1}$) 
of $\approx 10^6$ K plasma exceeds the escape speed, 
$v _{\rm esc}(r)=\sqrt{2G M_{\star}/r}$, at $r \gtrsim$ a few $R_{\star}$ 
in the red giant stars; 
the hot corona cannot be confined by the gravity any more in the atmospheres 
of the red giant stars.  
Therefore, the material flows out before heated up to coronal temperature.  

In addition, the thermal instability of the radiative cooling function 
(Landini \& Monsignori-Fossi 1990)
plays a role in the sudden decrease of temperature.  
Because of the thermal instability, magnetized hot  ($\gtrsim 10^6$K) bubbles  
intermittently exist in red giant winds, while most of the wind material 
consist of cool ($\lesssim 10^4$K) chromospheric gas (Suzuki 2007).  


\section{conclusions}
We have performed 1D MHD numerical simulations of solar and stellar winds 
from the photosphere. The low-frequency \Alfven waves are 
generated by the footpoint fluctuations of the magnetic field lines.  
We have treated the wave propagation and dissipation, and the heating and 
acceleration of the plasma in a self-consistent manner. Our simulation is 
the first simulation which treats the wind from the real surface 
(photosphere) to the (inner) heliosphere with the relevant physical processes. 

We have shown that the dissipation of the low-frequency \Alfven waves 
through the generation of the compressive waves (decay instability) and 
shocks (nonlinear steepening) is one of the solutions for the heating and 
acceleration of the plasma in the coronal holes.    
However, we should cautiously examine the validity of the 1-D MHD 
approximation we have adopted. 
There are other dissipation mechanisms due to the 
multidimensionality, such as turbulent cascade into the transverse direction 
(Goldreich \& Sridhar 1995; Oughton et al.2001)
and phase mixing (Heyvaerts \& Priest 1983). 
If \Alfven waves cascade to higher frequency, kinetic effects (e.g. 
Nariyuki \& Hada 2006) becomes important. 



We have also extended the solar wind simulations to red giant winds. 
With stellar evolution, the steady hot corona with temperature, 
$T \approx 10^6$ K, suddenly disappears because the surface gravity 
becomes small; hot plasma cannot be confined by the gravity.   
Thermal instability also generate intermittent magnetized hot bubbles 
in cool chromospheric winds. 

\begin{acknowledgements}
The author thanks the organizers of IAU247 for the nice conference.  
This work is supported in part by a Grant-in-Aid for Scientific
Research (19015004) from the Ministry of
Education, Culture, Sports, Science, and Technology of Japan.
\end{acknowledgements}


\end{document}